\documentclass[preprint,12pt]{elsarticle}

\usepackage[T1]{fontenc}
\usepackage[utf8]{inputenc}
\usepackage{lmodern}

\usepackage[none]{hyphenat}   
\usepackage{amsmath,amssymb}
\usepackage{graphicx}
\usepackage{booktabs}
\usepackage{siunitx}
\usepackage{natbib}
\usepackage{hyperref}

\allowdisplaybreaks

\journal{Environmental Challenges}

\biboptions{numbers,sort&compress}

\begin{document}

\begin{frontmatter}

\title{A Machine Learning--Based Surrogate EKMA Framework for Diagnosing Urban Ozone Formation Regimes: Evidence from Los Angeles}

\author{Sijie Zheng}
\ead{zhengsj0212@gmail.com}
\affiliation{
  organization={Department of Biostatistics},
  institution={University of California, Los Angeles},
  city={Los Angeles},
  country={USA}
}

\begin{abstract}
Surface ozone pollution remains a persistent challenge in many metropolitan regions worldwide,
as the nonlinear dependence of ozone formation on nitrogen oxides and volatile organic compounds (VOCs)
complicates the design of effective emission control strategies.
While chemical transport models provide mechanistic insights,
they rely on detailed emission inventories and are computationally expensive.

This study develops a machine learning--based surrogate framework inspired by the Empirical Kinetic Modeling Approach (EKMA).
Using hourly air quality observations from Los Angeles during 2024--2025,
a random forest model is trained to predict surface ozone concentrations based on precursor measurements
and spatiotemporal features, including site location and cyclic time encodings.
The model achieves strong predictive performance, with permutation importance highlighting
the dominant roles of diurnal temporal features and nitrogen dioxide,
along with additional contributions from carbon monoxide.

Building on the trained surrogate, EKMA-style sensitivity experiments are conducted by perturbing precursor concentrations
while holding other covariates fixed.
The results indicate that ozone formation in Los Angeles during the study period is predominantly VOC-limited.
Overall, the proposed framework offers an efficient and interpretable approach
for ozone regime diagnosis in data-rich urban environments.
\end{abstract}

\begin{keyword}
Surface ozone \sep Machine learning \sep Random forest \sep EKMA \sep Ozone formation regime \sep Urban air quality
\end{keyword}

\end{frontmatter}

\section{Introduction}

Surface ozone (O$_3$) is a key secondary air pollutant formed through photochemical reactions involving nitrogen oxides (NO$_x$) and volatile organic compounds (VOCs) under solar radiation \citep{wang2017ozone,jacob2009climate}. Elevated ozone concentrations pose serious risks to human health and ecosystems and remain difficult to control despite decades of regulatory effort, particularly in large metropolitan regions such as Los Angeles \citep{solomon2000comparison}.

The nonlinear response of ozone formation to its precursors is well documented in both field observations and modeling studies \citep{solomon2000comparison,tonnesen2000analysis}. Depending on chemical conditions, ozone production may be VOC-limited, NO$_x$-limited, or transitional, implying that precursor emission controls can lead to unintended outcomes. Such regime-dependent behavior has been extensively documented for the Los Angeles basin, which has long served as a benchmark region for urban photochemical smog research \citep{sillman2002sensitivity}. Accurate diagnosis of ozone formation regimes is therefore essential for effective mitigation strategies.

Chemical transport models (CTMs) have traditionally been used to assess ozone sensitivity through EKMA-style analyses \citep{dodge1977combined}. While mechanistically rich, CTMs require detailed emission inventories, chemical mechanisms, and substantial computational resources, and their results can be sensitive to uncertain inputs at urban scales \citep{tonnesen2000analysis}.

Recent advances in machine learning have enabled flexible modeling of air quality processes using observational data \citep{li2011application}. Random forest models, in particular, have demonstrated strong performance in capturing nonlinear relationships between air pollutants and meteorology and have been widely used for meteorological normalization and trend analysis \citep{grange2018random}. Moreover, machine learning models can serve as surrogates for complex physical systems, enabling efficient sensitivity analyses and policy-relevant scenario exploration \citep{carslaw2017detecting,yang2021covid}.

This study proposes a machine learning--based surrogate EKMA framework to diagnose ozone formation regimes without explicit chemical modeling. Los Angeles is selected as a case study due to its long history of photochemical smog and extensive monitoring network. Using data from 2024 to 2025, this study aims to characterize recent ozone dynamics, identify key drivers of ozone variability, and diagnose ozone formation regimes through an ML-based EKMA surrogate analysis.

\section{Data}

\subsection{Air quality observations}
Hourly air quality data from January 2024 to December 2025 were obtained from the
U.S. EPA Air Quality System (AQS) AirData repository, which provides regulatory
monitoring data across the United States.
Observations of surface ozone (O$_3$), nitrogen dioxide (NO$_2$), carbon monoxide (CO),
and fine particulate matter (PM$_{2.5}$) were collected for multiple monitoring sites
located within the Los Angeles metropolitan area.
Stations with insufficient temporal coverage or substantial data gaps were excluded
to ensure reliable descriptive analyses and model training.

The primary analysis focuses on hourly ozone concentrations (O$_3$--1h),
which are used for model estimation and EKMA-style sensitivity experiments.
Standard data quality checks were applied, including unit harmonization and the removal
of invalid or flagged observations following AQS reporting conventions.

\subsection{Meteorological variables}
Meteorological variability strongly influences ozone formation through its effects on photochemistry and atmospheric mixing.
In this study, high-frequency meteorological predictors (e.g., temperature, solar radiation, boundary-layer height)
are not explicitly included due to data availability and spatial matching constraints.
Instead, cyclic temporal encodings (hour of day, day of week, and month of year) are used to condition the analysis
on dominant diurnal and seasonal structure.
Consequently, all results should be interpreted as conditional statistical sensitivities under typical observed conditions,
rather than mechanistic responses from a chemical transport model.

\section{Methods}
Semiparametric approaches have also been used to analyze ozone-related time trends, including partially linear models that allow formal inference for smooth trend components \citep{ZhengSong2025}. While such models offer theoretical guarantees for trend estimation, they impose structural constraints that limit flexibility in capturing complex nonlinear interactions among multiple precursors. In contrast, the random forest surrogate adopted here provides greater modeling flexibility for high-dimensional and nonlinear ozone--precursor relationships, which is essential for EKMA-style sensitivity analysis.

\subsection{Missing data imputation}

Hourly air quality observations exhibit non-negligible missingness due to instrument downtime and site-specific reporting gaps.
To preserve multivariate relationships among pollutants, site location, and time, missing values were imputed using a $k$-nearest neighbors (KNN) procedure.

Let $\mathbf{z}_i \in \mathbb{R}^p$ denote the feature vector for record $i$, constructed from the predictors described in Section~\ref{sec:feature}.
All numeric features were standardized using training-set statistics:
$\tilde z_{ij} = (z_{ij} - \mu_j)/s_j$, where $(\mu_j, s_j)$ are the mean and standard deviation of feature $j$ computed on the training data.
For each record with missing entries, Euclidean distances were computed to all training records using the standardized feature vectors and the $k=5$ nearest neighbors were retained.
Each missing variable was imputed by the (unweighted) mean of that variable among the $k$ nearest neighbors.
For the held-out test set, standardization and neighbor search were performed using training-set statistics and the imputed training pool to avoid information leakage.


\subsection{Feature construction and temporal encoding}\label{sec:feature}

To capture the strong periodicity in surface ozone, hour of day, month of year, and day of week were encoded using continuous cyclic representations.
Specifically, hour of day was transformed into sine--cosine pairs as
\[
\text{hour\_sin} = \sin\!\left(2\pi \cdot \frac{\text{hour}}{24}\right),
\qquad
\text{hour\_cos} = \cos\!\left(2\pi \cdot \frac{\text{hour}}{24}\right),
\]
with analogous transformations applied to month and day-of-week.

The final predictor set included NO$_2$, CO,
PM$_{2.5}$, spatial coordinates,
and the cyclic temporal features described above.

\subsection{Random forest modeling of surface ozone}

Surface ozone concentrations were modeled using a random forest (RF) regression surrogate.
Random forests aggregate an ensemble of decision trees trained on bootstrap samples, while randomly subsampling candidate features at each split.
This design yields strong predictive accuracy in nonlinear settings and provides a convenient surrogate for sensitivity experiments.

Let $\mathbf{X}_t$ denote the predictor vector at time $t$ after imputation and feature construction.
The RF surrogate estimates the conditional mean
\[
\hat{O}_3(t) = f_{\text{RF}}(\mathbf{X}_t),
\]
where $f_{\text{RF}}(\cdot)$ is the ensemble prediction.
Model training used data from 2024 and evaluation used an independent test set from 2025 (temporal split) to prevent leakage across time.

In implementation, the RF model was fit using the \texttt{ranger} package with 500 trees (\texttt{num.trees}=500) and a fixed random seed for reproducibility.
Unless stated otherwise, remaining hyperparameters followed \texttt{ranger} defaults (e.g., $m_{\text{try}}=\lfloor\sqrt{p}\rfloor$ for regression).
Predictive performance was summarized by the coefficient of determination ($R^2$) and the root mean squared error (RMSE) on the 2025 test set.

To quantify the contribution of each predictor, permutation importance was computed on the test set by measuring the increase in RMSE induced by random permutation of each feature, with all other features held fixed.
This model-agnostic measure captures the marginal importance of each feature for out-of-sample prediction.

\subsection{RF-based surrogate EKMA sensitivity analysis}

To diagnose ozone formation regimes, we conducted an Empirical Kinetic Modeling Approach (EKMA)--inspired sensitivity analysis using the trained RF model as a statistical surrogate for the ozone response surface.
Traditional EKMA diagrams evaluate ozone responses to joint NO$_x$ and VOC emission changes under fixed spatiotemporal covariates within the observational surrogate.
Here, we perturb observed precursor concentrations within a data-driven surrogate, conditioning on realistic spatiotemporal covariates.
Throughout this study, the term ``ozone formation regime'' refers to a conditional sensitivity regime inferred within the observational surrogate framework, rather than a mechanistic regime defined by emission-rate perturbations in chemical transport models.

Nitrogen dioxide (NO$_2$) was used as an observational proxy for NO$_x$,
and carbon monoxide (CO) was used as a proxy for VOC-related emissions.
Although CO is not a reactive VOC, it is commonly employed as an observational proxy
for combustion-related VOC emissions due to shared sources and strong co-variability
in urban environments.

We defined a baseline sample set $\mathcal{B}$ intended to represent typical
high-ozone conditions in Los Angeles.
Specifically, $\mathcal{B}$ consisted of observations from year 2024,
summer months (June--September), and afternoon hours (12:00--17:00 local time).
All other covariates, including spatial coordinates and cyclic time features,
were held fixed at their observed baseline values.

For scaling factors $\alpha,\beta \in [0.5,1.5]$, precursor concentrations were perturbed as
\[
\text{NO}_2^{(\alpha)} = \alpha \cdot \text{NO}_2, \qquad
\text{CO}^{(\beta)} = \beta \cdot \text{CO},
\]
and the random forest surrogate was evaluated on the modified baseline covariates.
The scaling range was chosen to represent moderate perturbations around observed conditions
while avoiding extrapolation far beyond the data support.

For each $(\alpha,\beta)$ pair, predicted ozone values were averaged over all samples
in $\mathcal{B}$ to obtain an EKMA-like response surface
\[
\bar O_3(\alpha,\beta) =
\frac{1}{|\mathcal{B}|}\sum_{i\in\mathcal{B}}
f_{\mathrm{RF}}\!\left(\mathbf{X}_i^{(\alpha,\beta)}\right).
\]
Averaging over $\mathcal{B}$ yields a conditional mean response surface that
marginalizes over realistic within-baseline variability rather than representing
a single deterministic state.

Ozone formation regimes were interpreted based on relative sensitivity:
if $\bar O_3$ decreases more strongly with reductions in CO (smaller $\beta$)
than with comparable reductions in NO$_2$ (smaller $\alpha$),
this behavior is indicative of VOC-limited conditions;
the opposite pattern suggests NO$_x$-limited behavior.

\section{Results}


\subsection{Temporal characteristics of surface ozone}

\begin{figure}[!htbp]
\centering
\includegraphics[width=0.7\linewidth]{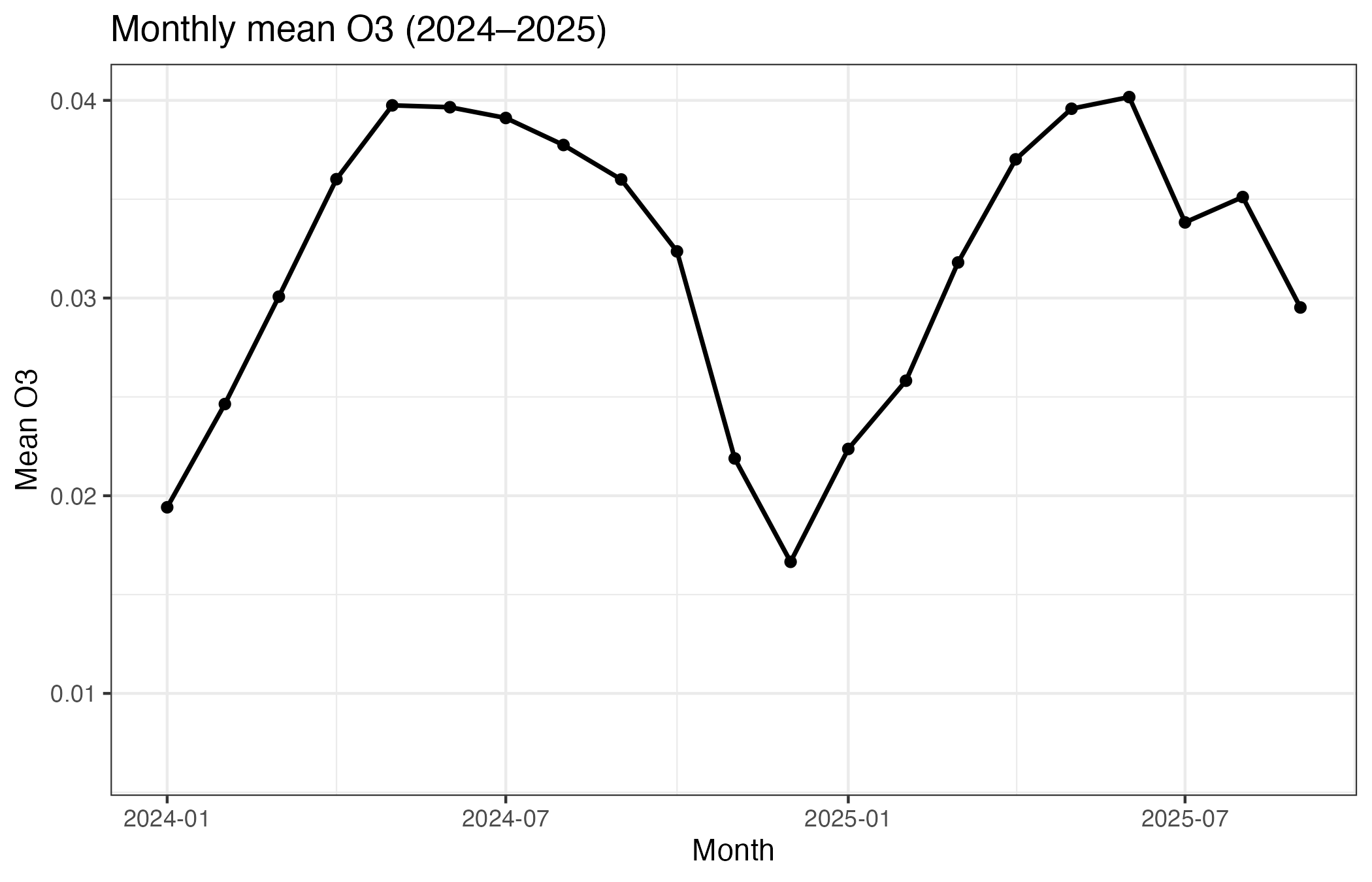}
\caption{
Monthly mean O$_3$ in Los Angeles during 2024--2025.
Ozone concentrations exhibit pronounced seasonality, with elevated levels in late spring and summer
and substantially lower concentrations during winter months.
}
\label{fig:o3_time_a}
\end{figure}

\begin{figure}[!htbp]
\centering
\includegraphics[width=0.7\linewidth]{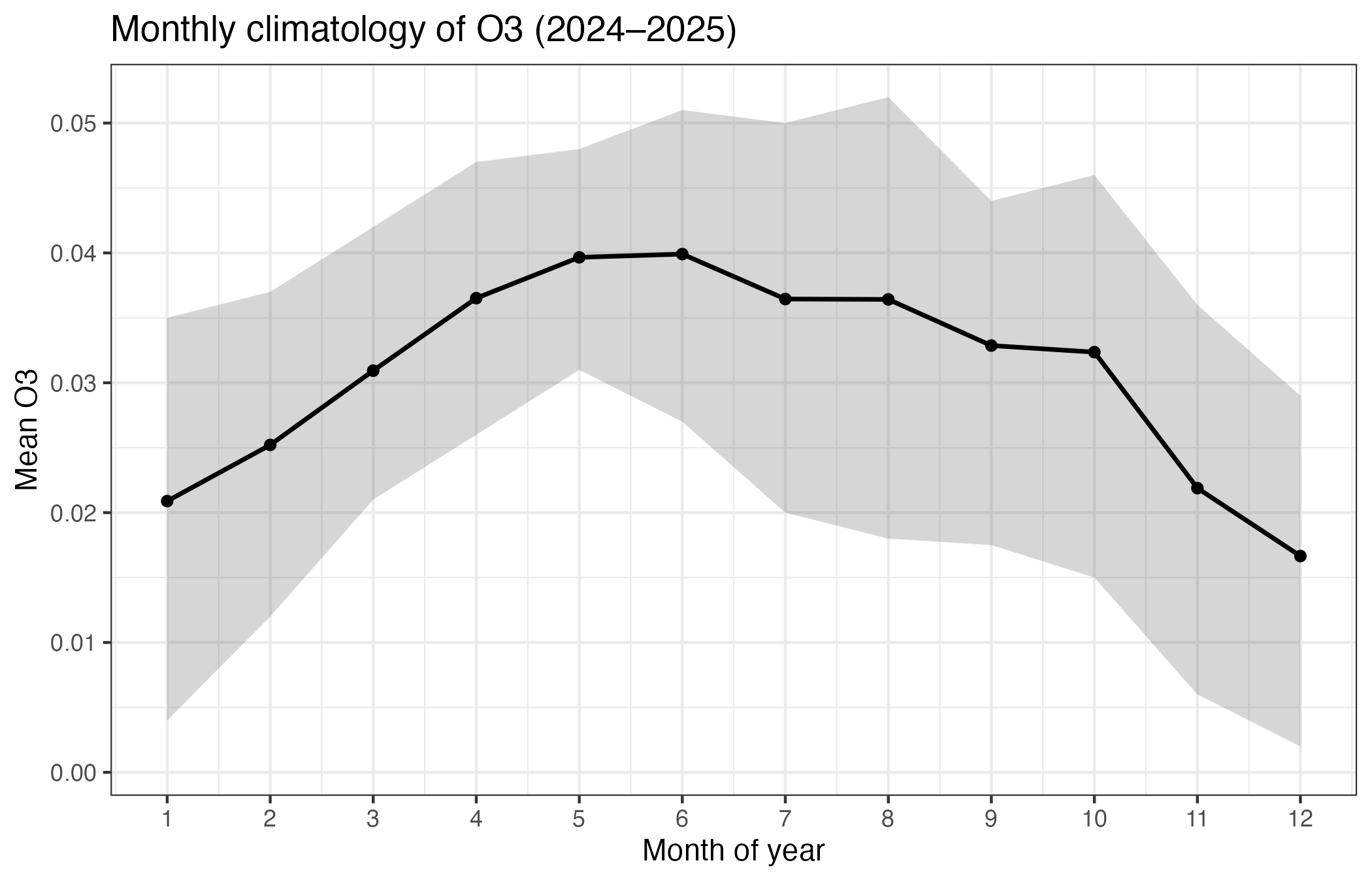}
\caption{
Monthly climatology of O$_3$ in Los Angeles aggregated across 2024--2025.
The solid line indicates the mean concentration, while the shaded band represents the interquartile range,
highlighting increased variability during the warm season.
}
\label{fig:o3_time_b}
\end{figure}

\begin{figure}[!htbp]
\centering
\includegraphics[width=0.7\linewidth]{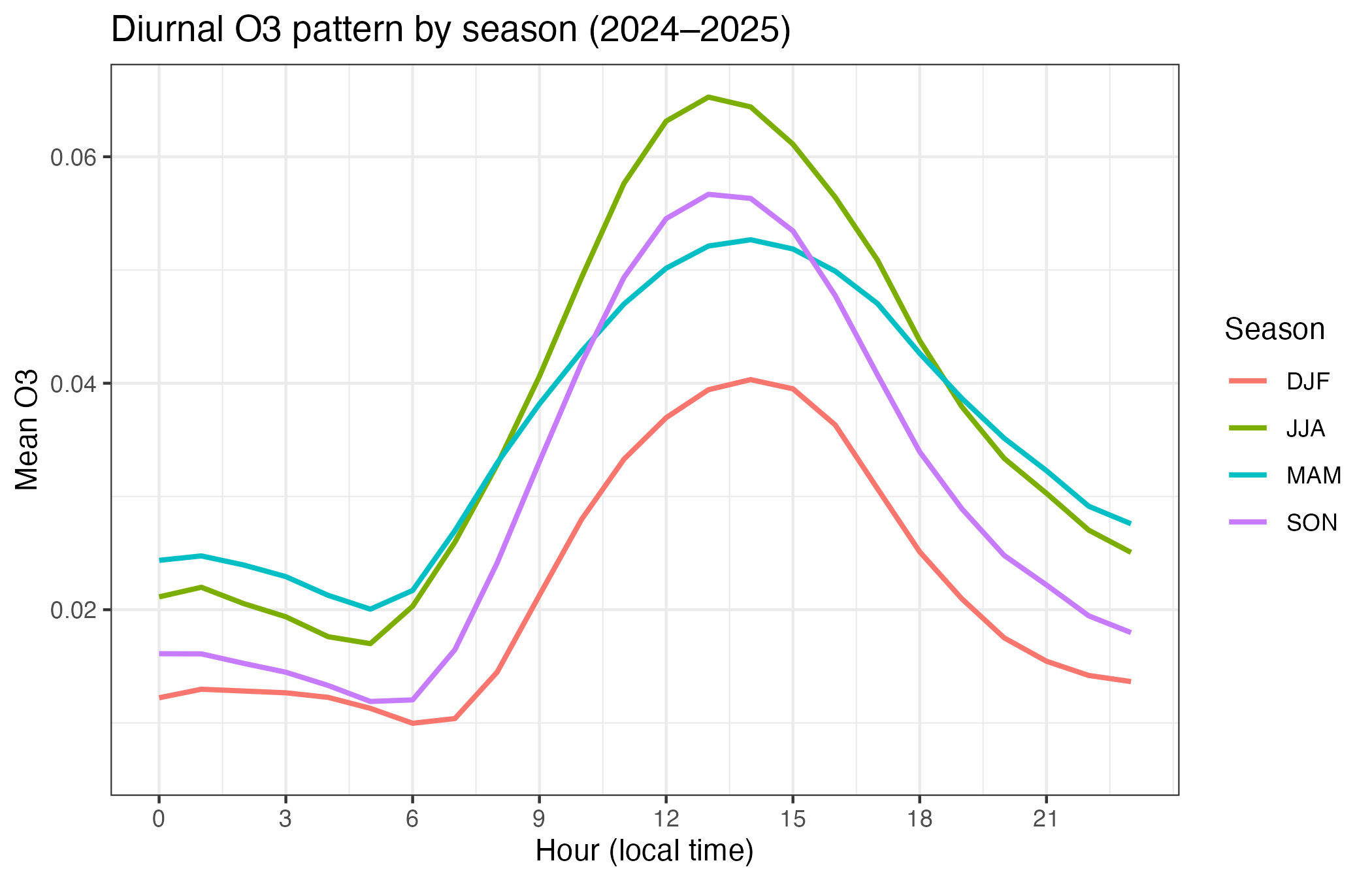}
\caption{
Mean diurnal cycle of O$_3$ by season (DJF, MAM, JJA, SON) in Los Angeles during 2024--2025.
Ozone concentrations increase rapidly after sunrise, peak in the early afternoon,
and decline during the evening, with the largest diurnal amplitude observed in summer (JJA).
}
\label{fig:o3_time_c}
\end{figure}

Surface ozone in Los Angeles exhibits pronounced variability across monthly,
seasonal, and diurnal time scales
(Figure~\ref{fig:o3_time_a}--\ref{fig:o3_time_c}).
Monthly mean ozone concentrations
(Figure~\ref{fig:o3_time_a})
show a clear seasonal pattern, with elevated levels during late spring and summer
and substantially lower concentrations in winter.
This seasonal cycle reflects the combined effects of enhanced solar radiation,
higher temperatures, and more favorable photochemical conditions during the warm season.

To further characterize seasonal variability, Figure~\ref{fig:o3_time_b}
presents the monthly climatology.
Mean ozone concentrations peak in late spring and summer (approximately May--August),
while the interquartile range widens noticeably during the warm season,
indicating increased variability under high-temperature and high-radiation conditions.
In contrast, winter months are characterized by both lower mean ozone levels
and reduced variability.

At the diurnal scale, surface ozone displays a clear single-peak pattern
(Figure~\ref{fig:o3_time_c}).
Ozone concentrations are lowest during the early morning hours,
increase rapidly after sunrise, and reach maximum values in the early afternoon.
The diurnal amplitude is strongly season-dependent, with the largest daytime
ozone enhancement observed in summer (JJA), followed by spring (MAM) and autumn (SON),
and the weakest diurnal cycle occurring in winter (DJF).
This behavior is consistent with photochemical ozone production driven by
solar radiation and temperature, as well as nighttime titration by nitrogen oxides.

\subsection{Weekly patterns and the weekend effect}

\begin{figure}[!htbp]
\centering
\includegraphics[width=0.7\linewidth]{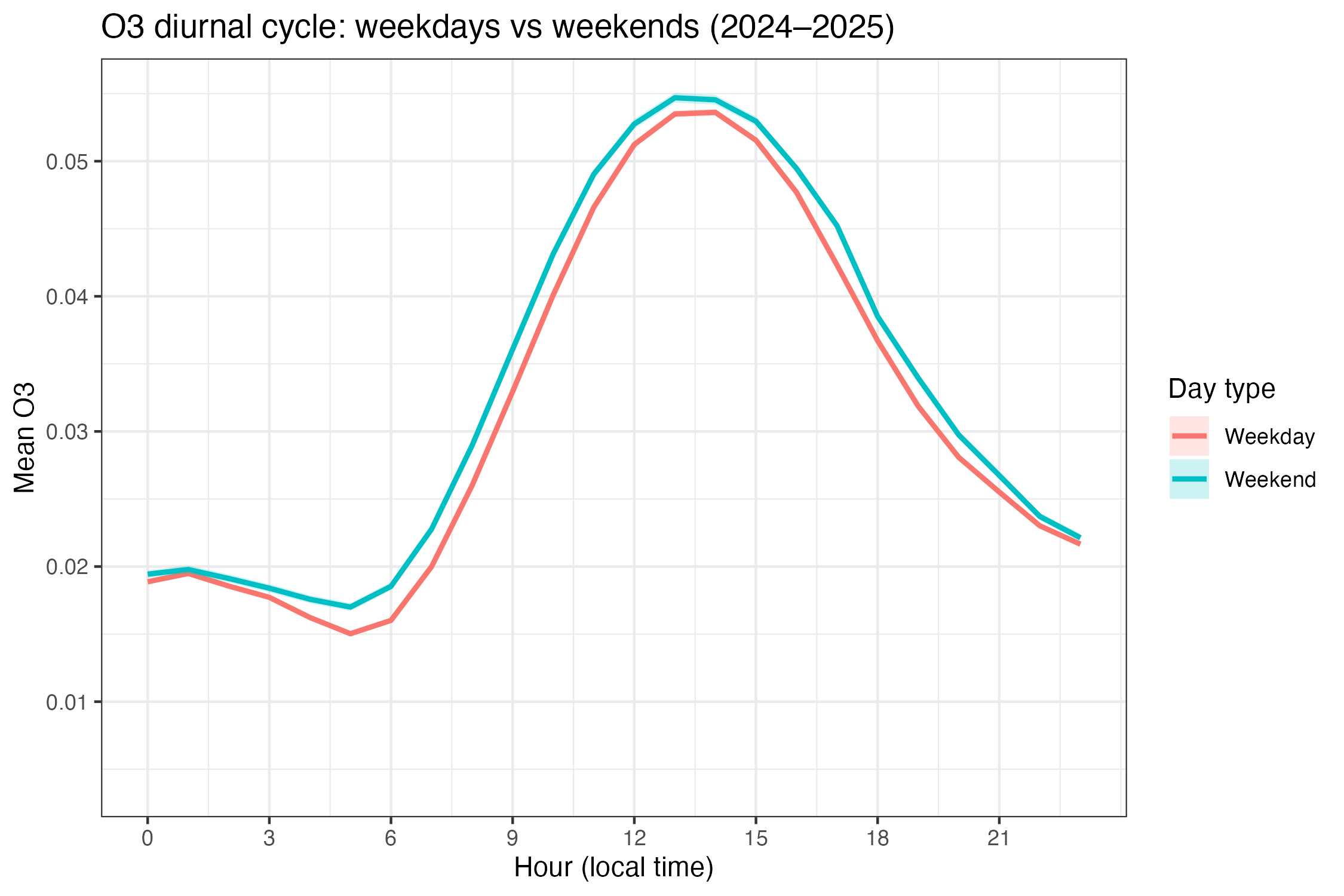}
\caption{
Diurnal ozone cycles averaged over weekdays and weekends in Los Angeles during 2024--2025.
Daytime ozone concentrations are systematically higher on weekends,
consistent with the ozone weekend effect and indicative of nonlinear ozone--precursor interactions.
}
\label{fig:o3_weekend}
\end{figure}

Figure~\ref{fig:o3_weekend} compares diurnal ozone cycles averaged over weekdays
and weekends.
A clear weekend effect is observed, with systematically higher daytime ozone
concentrations on weekends than on weekdays.

The weekend--weekday difference is most pronounced during the afternoon hours,
coinciding with the period of maximum photochemical ozone production,
while nighttime differences remain relatively small.
This pattern is consistent with reduced NO$_x$ emissions on weekends,
which weaken ozone titration and enhance net ozone accumulation,
particularly under nonlinear ozone--precursor chemistry.


\subsection{Random forest model performance}
\begin{figure}[!htbp]
\centering
\includegraphics[width=0.7\linewidth]{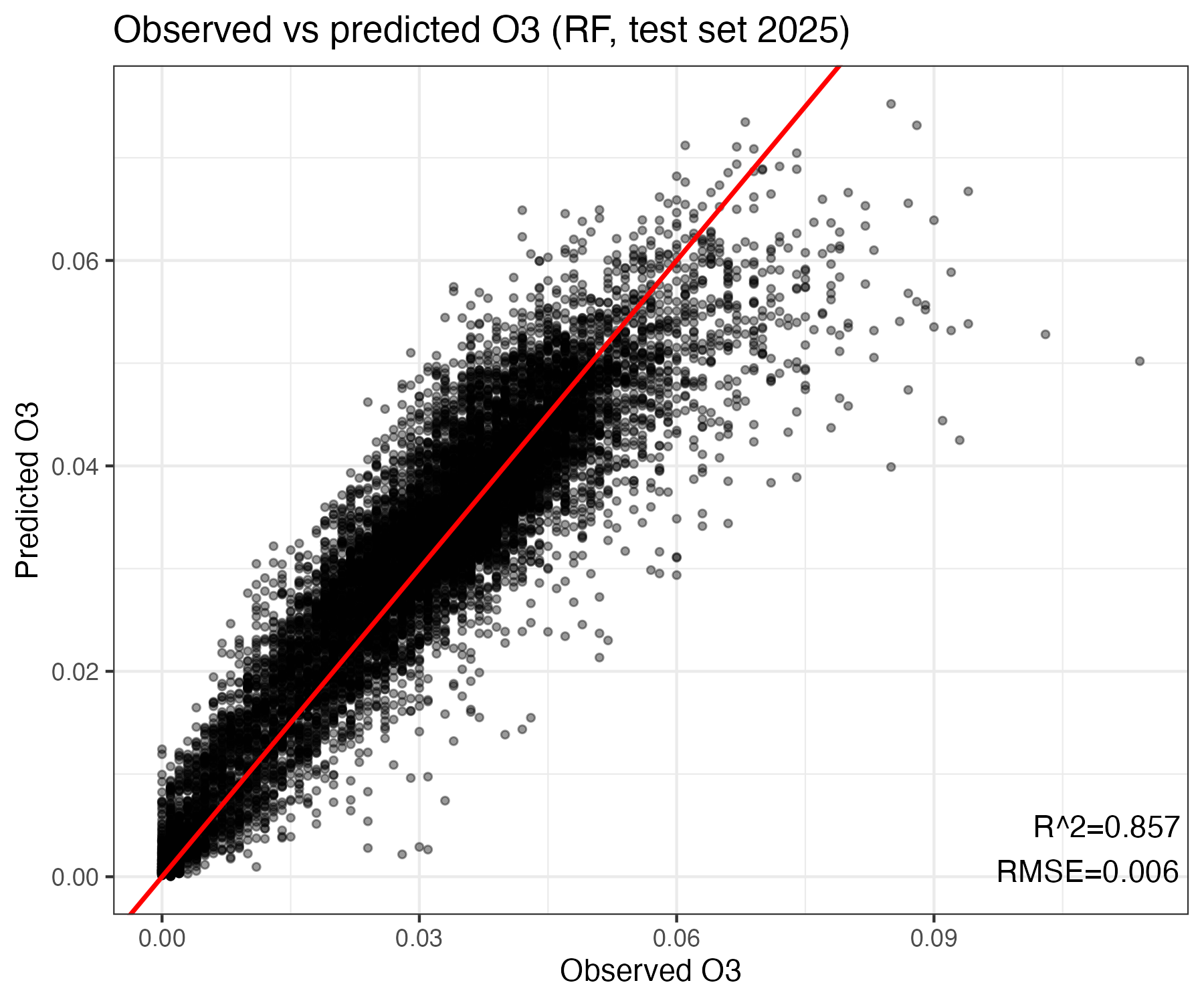}
\caption{
Observed versus predicted surface ozone (O$_3$) for the random forest surrogate on the 2025 test set.
The red line indicates the 1:1 reference. Reported metrics are $R^2=0.857$ and $\mathrm{RMSE}=0.006$.
}
\label{fig:rf_perf}
\end{figure}

Figure~\ref{fig:rf_perf} evaluates the predictive performance of the random forest surrogate
using an independent test set from 2025.
The model achieves strong out-of-sample accuracy ($R^2=0.857$; $\mathrm{RMSE}=0.006$),
and predicted ozone concentrations closely follow the 1:1 reference line over a broad range of observed values.

Performance is particularly robust for moderate ozone concentrations, which are most relevant for regime diagnosis.
At the upper tail, the model shows a mild tendency to smooth extreme values, which is common in data-driven regressors
when extreme episodes are sparse.
Overall, the RF surrogate provides a reliable basis for subsequent sensitivity experiments.

\subsection{Key drivers of ozone variability}
\begin{figure}[!htbp]
\centering
\includegraphics[width=0.7\linewidth]{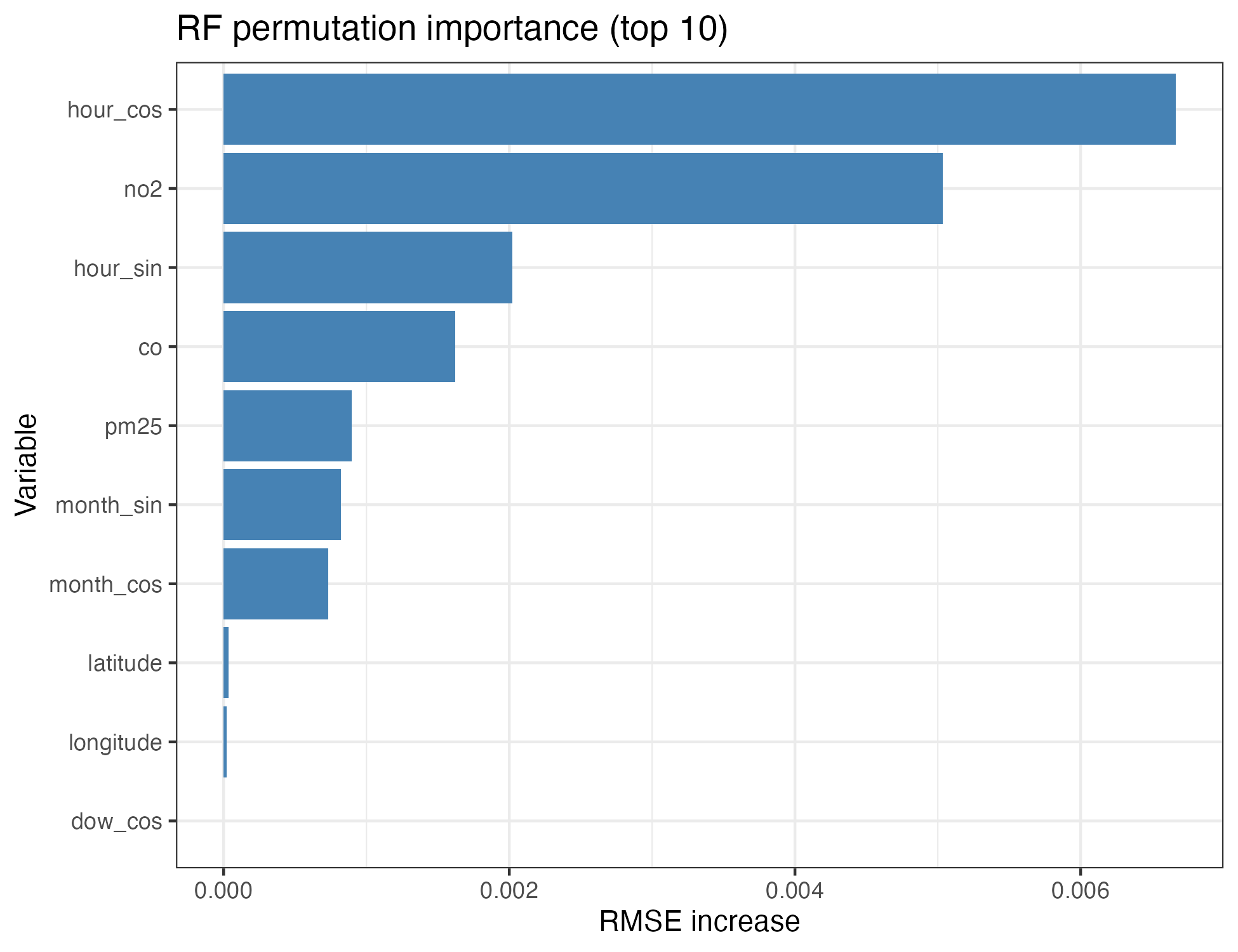}
\caption{
Permutation importance (top 10 predictors) for the random forest ozone model,
measured as the increase in test-set RMSE after permuting each feature.
Cyclic hour-of-day features (hour\_sin, hour\_cos) and NO$_2$ rank among the most influential predictors.
}
\label{fig:rf_importance}
\end{figure}

Figure~\ref{fig:rf_importance} shows permutation importance (top 10 predictors) for the RF model.
Cyclic time features for hour-of-day (hour\_sin, hour\_cos) rank among the most influential predictors,
highlighting the dominant role of the diurnal photochemical cycle.
NO$_2$ also exhibits high importance, consistent with its dual role as an ozone precursor and as a nighttime titrant.

CO, used here as a proxy for VOC-related emissions, contributes substantially to predictive skill,
supporting its relevance for surrogate EKMA perturbations.
PM$_{2.5}$ and seasonal features (month\_sin/month\_cos) show additional influence,
reflecting seasonal variability and correlated atmospheric processes.

\subsection{ML-based surrogate EKMA response surfaces}
\begin{figure}[!htbp]
\centering
\includegraphics[width=0.7\linewidth]{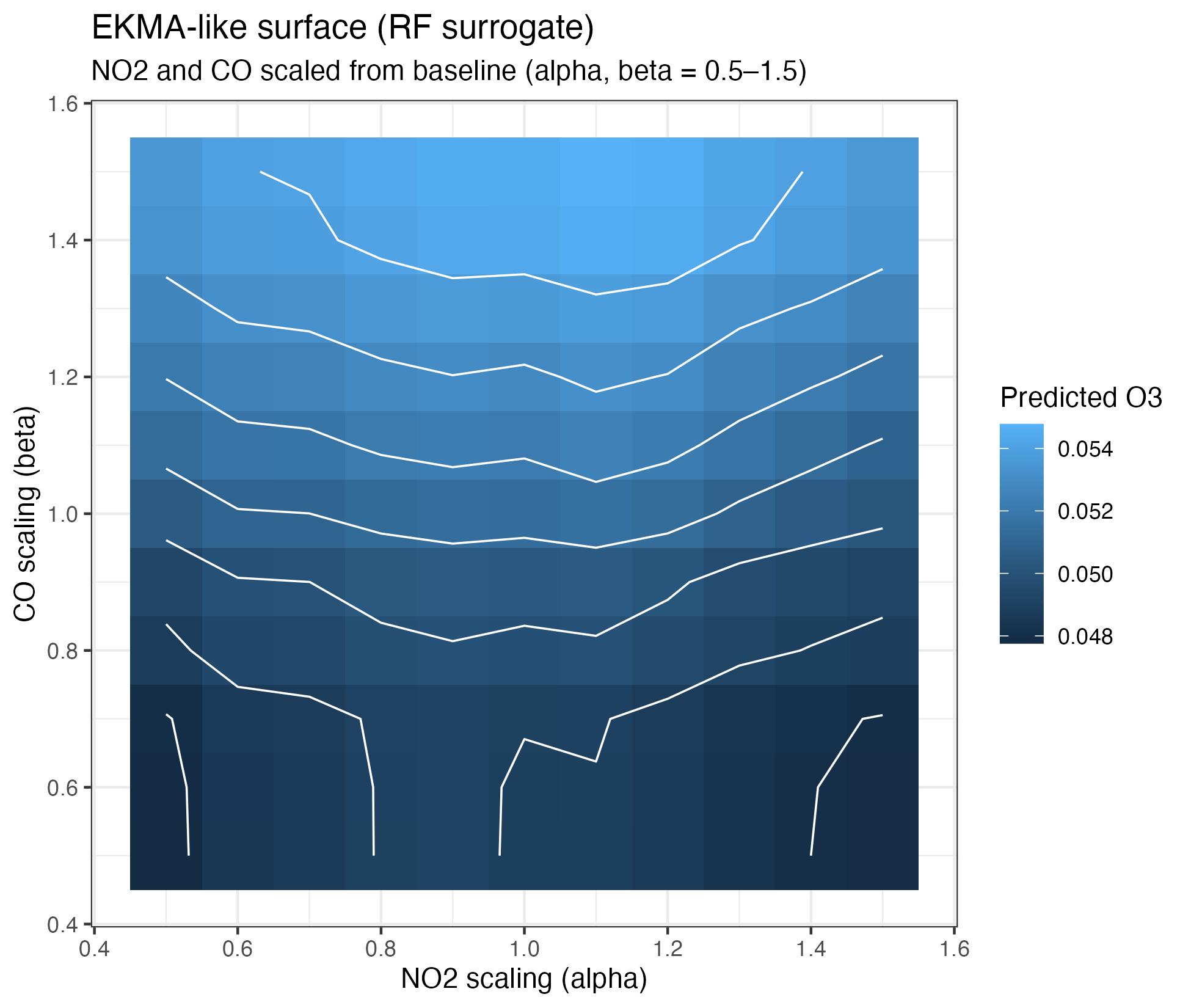}
\caption{
EKMA-like ozone response surface derived from the random forest surrogate.
NO$_2$ and CO are scaled from baseline by factors $\alpha$ and $\beta$ in $[0.5,1.5]$, respectively,
while other covariates are held fixed. White curves denote predicted ozone isopleths.
}
\label{fig:ekma_2d}
\end{figure}

Figure~\ref{fig:ekma_2d} presents an EKMA-like response surface derived from the RF surrogate by jointly scaling NO$_2$
and CO concentrations over $\alpha,\beta\in[0.5,1.5]$ while holding other covariates fixed.
Predicted ozone exhibits stronger sensitivity to CO scaling than to comparable NO$_2$ scaling across much of the grid,
which indicates a predominantly VOC-sensitive regime within the observational surrogate framework.

\begin{figure}[!htbp]
\centering
\includegraphics[width=0.7\linewidth]{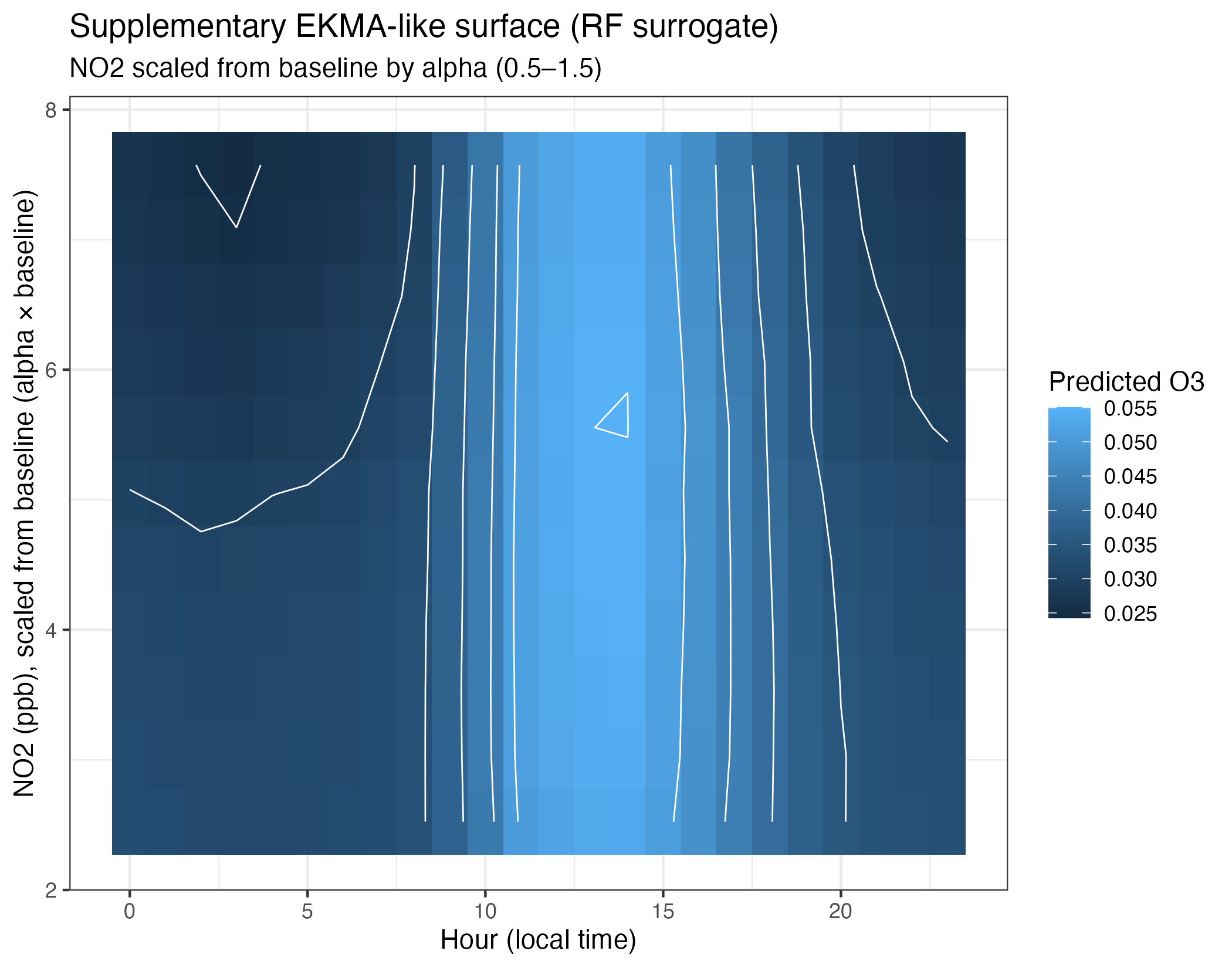}
\caption{
Supplementary EKMA-like response surface showing predicted ozone as a function of local hour and scaled NO$_2$
($\alpha\in[0.5,1.5]$), based on the random forest surrogate model.
}
\label{fig:ekma_hour_no2}
\end{figure}

To further illustrate ozone sensitivity to NO$_2$ under different hours of day, Figure~\ref{fig:ekma_hour_no2}
provides a supplementary response surface with hour on the x-axis and scaled NO$_2$ on the y-axis.
Together, these results support the conclusion that ozone formation during 2024--2025 in Los Angeles is predominantly
VOC-limited under the observational surrogate framework.

\section{Discussion}

\subsection{Interpretation of ozone formation regimes}
The results from the ML-based surrogate EKMA analysis consistently indicate that surface ozone formation in Los Angeles during 2024--2025 is predominantly VOC-limited. This finding is supported by multiple lines of evidence, including the strong weekend effect, the high importance of nitrogen dioxide and cyclic temporal features (diurnal and seasonal encodings) in the random forest model, and the greater ozone sensitivity to VOC perturbations relative to NO$_2$ perturbations.

The persistence of this regime suggests that reductions in NO$_x$ emissions alone
may be insufficient or even counterproductive for ozone mitigation,
particularly during high-temperature and high-radiation conditions. Instead, coordinated control of VOC emissions is likely to yield more effective ozone reductions. These conclusions are consistent with established photochemical theory and prior observational studies in photochemically active urban regions.

\subsection{Advantages of the ML-based surrogate EKMA framework}
A key contribution of this study lies in the integration of EKMA concepts with a machine learning surrogate model trained on observational data. Traditional EKMA analyses rely on explicit chemical transport models, which require detailed emission inventories, chemical mechanisms, and substantial computational resources. In contrast, the proposed framework directly leverages observed relationships between ozone, its precursors, and meteorological conditions.

By using a random forest model as a surrogate for the ozone response surface, the framework captures nonlinear interactions and regime-dependent sensitivities without explicit specification of reaction pathways. This approach enables efficient sensitivity experiments under fixed spatiotemporal covariates within the observational surrogate and provides interpretable diagnostics that closely resemble classical EKMA diagrams. As a result, the surrogate EKMA framework offers a practical alternative for regime diagnosis in data-rich urban environments where emission inventories or chemical modeling resources are limited.

\subsection{Policy implications}
The identification of a VOC-limited ozone formation regime has direct implications for air quality management in Los Angeles and similar metropolitan regions. Emission control strategies that prioritize VOC reductions, such as controls on solvent use, fuel evaporation, industrial processes, and mobile-source VOC emissions, are likely to be more effective in reducing peak ozone levels than strategies focused solely on NO$_x$ reductions.

Moreover, the ML-based surrogate EKMA framework can be readily updated as new observational data become available, allowing policymakers to track potential regime shifts over time. This flexibility is particularly valuable under changing climate conditions, where increasing temperatures and heatwave frequency may further amplify photochemical ozone production.

\subsection{Limitations and future work}

Several limitations should be acknowledged.
First, the main analysis does not include explicit meteorological predictors
(e.g., temperature, boundary-layer height, or solar radiation) at the hourly resolution.
Although cyclic temporal features partially proxy for meteorological influences
through seasonal and diurnal structure, they cannot capture day-to-day synoptic variability
or extreme heat events.
As a result, the random forest surrogate and EKMA-like perturbations
should be interpreted as conditional statistical sensitivities within the observed covariate support,
and regime classification may be less reliable outside typical summer-afternoon conditions.

Second, the surrogate EKMA analysis relies on observed precursor concentrations
rather than explicit emission rates, which may introduce uncertainty when interpreting
control scenarios in terms of specific source categories.
In addition, while random forest models provide strong predictive performance,
they remain statistical approximations and cannot fully replace
process-based chemical transport models for source attribution
or long-range transport analysis.

Future work could address these limitations by integrating gridded meteorological
reanalysis matched by time and location, incorporating additional predictors
such as speciated VOC measurements or satellite-derived indicators,
and exploring machine learning models with improved interpretability.
A further extension is to treat ozone concentration trajectories
as functional data observed over time.
Recent advances in functional data inference provide a rigorous framework
\citep{HuangZhengYang2022,ZhengHuangYang2025,ZhengMengZhou2025},
which could be combined with surrogate-based EKMA diagnostics
to enable more principled regime classification under temporal dependence.

\section{Conclusions}
This study presents a machine learning--based surrogate EKMA framework for diagnosing urban ozone formation regimes using observational data. Applying the framework to Los Angeles from 2024 to 2025 reveals strong seasonal, diurnal, and weekly ozone patterns and indicates that recent ozone formation is predominantly VOC-limited.

By integrating random forest modeling with EKMA-inspired sensitivity analysis, the proposed approach provides an efficient and interpretable alternative to traditional chemical transport modeling for regime diagnosis. The results underscore the importance of VOC emission controls for effective ozone mitigation and highlight the potential of data-driven methods to support air quality management in complex urban environments.

\section*{Data and code availability}
All air quality observations used in this study were obtained from the U.S. EPA AQS AirData repository and are publicly available. The analysis code implementing data preprocessing, random forest modeling, and surrogate EKMA sensitivity analysis is publicly available at \url{https://github.com/zhengsj0212/surrogate-ekma-la-ozone}. The repository contains scripts sufficient to reproduce all figures and results presented in this paper using the archived AQS datasets. 

\section*{Figures}

\newpage

\bibliographystyle{elsarticle-harv}
\bibliography{ref}

@article{wang2017ozone,
  title={Ozone pollution in China: A review of concentrations, meteorological influences, chemical precursors, and effects},
  author={Wang, Tao and Xue, Likun and Brimblecombe, Peter and Lam, Yun Fat and Li, Li and Zhang, Li},
  journal={Science of the Total Environment},
  volume={575},
  pages={1582--1596},
  year={2017},
  publisher={Elsevier}
}

@article{solomon2000comparison,
  title={Comparison of scientific findings from major ozone field studies in North America and Europe},
  author={Solomon, Paul and Cowling, Ellis and Hidy, George and Furiness, Cari},
  journal={Atmospheric Environment},
  volume={34},
  number={12-14},
  pages={1885--1920},
  year={2000},
  publisher={Elsevier}
}

@inproceedings{dodge1977combined,
  title={Combined use of modeling techniques and smog chamber data to derive ozone-precursor relationships},
  author={Dodge, MC},
  booktitle={International conference on photochemical oxidant pollution and its control: Proceedings},
  volume={2},
  pages={881--889},
  year={1977}
}

@article{tonnesen2000analysis,
  title={Analysis of radical propagation efficiency to assess ozone sensitivity to hydrocarbons and NO x: 1. Local indicators of instantaneous odd oxygen production sensitivity},
  author={Tonnesen, Gail S and Dennis, Robin L},
  journal={Journal of Geophysical Research: Atmospheres},
  volume={105},
  number={D7},
  pages={9213--9225},
  year={2000},
  publisher={Wiley Online Library}
}

@article{li2011application,
  title={Application of machine learning methods to spatial interpolation of environmental variables},
  author={Li, Jin and Heap, Andrew D and Potter, Anna and Daniell, James J},
  journal={Environmental Modelling \& Software},
  volume={26},
  number={12},
  pages={1647--1659},
  year={2011},
  publisher={Elsevier}
}

@article{grange2018random,
  title={Random forest meteorological normalisation models for Swiss PM 10 trend analysis},
  author={Grange, Stuart K and Carslaw, David C and Lewis, Alastair C and Boleti, Eirini and Hueglin, Christoph},
  journal={Atmospheric Chemistry and Physics},
  volume={18},
  number={9},
  pages={6223--6239},
  year={2018},
  publisher={Copernicus GmbH}
}

@article{carslaw2017detecting,
  title={Detecting and characterising small changes in urban nitrogen dioxide concentrations},
  author={Carslaw, David C and Carslaw, Nicola},
  journal={Atmospheric Environment},
  volume={41},
  number={22},
  pages={4723--4733},
  year={2007},
  publisher={Elsevier}
}

@article{yang2021covid,
  title={From COVID-19 to future electrification: Assessing traffic impacts on air quality by a machine-learning model},
  author={Yang, Jiani and Wen, Yifan and Wang, Yuan and Zhang, Shaojun and Pinto, Joseph P and Pennington, Elyse A and Wang, Zhou and Wu, Ye and Sander, Stanley P and Jiang, Jonathan H and others},
  journal={Proceedings of the National Academy of Sciences},
  volume={118},
  number={26},
  pages={e2102705118},
  year={2021},
  publisher={National Academy of Sciences}
}

@article{jacob2009climate,
  title={Effect of climate change on air quality},
  author={Jacob, Daniel J and Winner, Darrell A},
  journal={Atmospheric environment},
  volume={43},
  number={1},
  pages={51--63},
  year={2009},
  publisher={Elsevier}
}

@article{sillman2002sensitivity,
  title={The sensitivity of ozone to nitrogen oxides and hydrocarbons in regional ozone episodes},
  author={Sillman, Sanford and Logan, Jennifer A and Wofsy, Steven C},
  journal={Journal of Geophysical Research: Atmospheres},
  volume={95},
  number={D2},
  pages={1837--1851},
  year={1990},
  publisher={Wiley Online Library}
}

@article{HuangZhengYang2022,
  title={Inference for dependent error functional data with application to event-related potentials},
  author={Huang, Kun and Zheng, Sijie and Yang, Lijian},
  journal={TEST},
  volume={31},
  number={4},
  pages={1100--1120},
  year={2022},
  publisher={Springer}
}

@article{ZhengSong2025,
  title={Inference for trend functions in partially linear models},
  author={Zheng, Sijie and Song, Xiaojun},
  journal={Journal of Statistical Planning and Inference},
  pages={106338},
  year={2025},
  publisher={Elsevier}
}

@article{ZhengHuangYang2025,
  title={Inference for dependent error functional data: Covariance function},
  author={Zheng, Sijie and Huang, Kun and Yang, Lijian},
  journal={Electronic Journal of Statistics},
  volume={19},
  number={2},
  pages={5216--5248},
  year={2025},
  publisher={The Institute of Mathematical Statistics and the Bernoulli Society}
}

@article{ZhengMengZhou2025,
  title   = {Efficient Covariance Estimation for Sparsified Functional Data},
  author  = {Zheng, Sijie and Meng, Fandong and Zhou, Jie},
  journal = {arXiv preprint arXiv:2511.18237},
  year    = {2025}
}

\end{document}